\documentclass[conference]{IEEEtran}
\usepackage{cite}
\usepackage{amsmath,amssymb}
\usepackage{booktabs}
\usepackage{multirow}
\usepackage{graphicx}
\usepackage{dblfloatfix}
\usepackage{siunitx}
\usepackage[export]{adjustbox}
\usepackage{url}
\usepackage{enumitem}
\usepackage{pgfplots}
\usetikzlibrary{arrows.meta,positioning}
\usepgfplotslibrary{groupplots}
\pgfplotsset{compat=1.17}
\usepackage{xcolor}

\pgfplotsset{compat=1.17}
\usepackage{comment}
\begin{document}

\title{When Simple Is Enough: Resource-Aware Heuristic versus DDQN for O-RAN Mobility Robustness}
\author{%
\IEEEauthorblockN{
Ali Deylamy\IEEEauthorrefmark{1}, and Nima Afraz\IEEEauthorrefmark{1}}
\IEEEauthorblockA{
\IEEEauthorrefmark{1}School of Computer Science,
University College Dublin, Dublin, Ireland
}

}

\maketitle

\begin{abstract}
Open Radio Access Network (O-RAN) enables mobility-management xApps, but a
learned controller is not necessarily the most effective or efficient handover
policy. This paper compares standard Event A3, a resource- and mobility-aware
heuristic, and a Double Deep Q-Network (DDQN) through a common xApp-oriented
interface and paired system-level simulation. The heuristic combines radio
quality, target-cell resources, UE mobility, and service class, with weights
optimized offline using NSGA-II, while DDQN-Multi uses temporal radio and
resource observations. All controllers are evaluated on identical UE
trajectories, traffic assignments, and ray-traced radio inputs under Base,
High-Mobility, and High-UE scenarios. Performance is assessed using service,
handover, resource, energy, and fairness KPIs, complemented by
CRITIC--TOPSIS ranking and complexity analysis. Results show that the heuristic
achieves the lowest outage in all three scenarios and ranks first in the
CRITIC--TOPSIS comparison, while retaining linear online complexity and
interpretable decisions. DDQN-Multi improves target feasibility under
congestion but requires offline training and produces more handovers and
ping-pong events. The results indicate that lightweight resource-aware control can provide a favourable performance-complexity trade-off relative to DDQN under the considered scenarios.
\end{abstract}


\section{Introduction}
\label{sec:introduction}
Reliable mobility management is essential for maintaining service continuity in
fifth-generation (5G) networks. As a user equipment (UE) moves, its serving cell
may become unsuitable because of propagation loss, interference, mobility, or
resource congestion. A handover controller must therefore select a suitable
target in time to avoid service degradation while limiting unnecessary
handovers, interruption, signalling overhead, admission failures, and
ping-pong events \cite{haghrah2023handover_nr}. These objectives are often
conflicting: aggressive policies may improve responsiveness but increase
handover activity, whereas conservative policies may leave UEs connected to
poorly performing cells.
3GPP Event A3 triggers handover reporting when a neighbouring-cell measurement
exceeds that of the serving cell by a configured offset for a specified
time-to-trigger (TTT) \cite{3gpp_ts_38331}. Although simple and interpretable,
its radio-based formulation does not explicitly account for target-cell
resources, UE mobility, or service conditions. Multi-criteria heuristics can
incorporate additional information such as RSRP, RSRQ, SINR, cell load,
available PRBs, and UE motion \cite{palas2021multicriteria}, but their
performance depends on the selected weights, thresholds, and decision rules.
Reinforcement learning provides an alternative by treating handover as a
sequential decision problem and accounting for longer-term service and mobility
outcomes. DDQN-based approaches are widely used for mobility management, but
their training requirements, computational complexity, and limited
interpretability motivate comparison with simpler resource-aware controllers.
A meaningful comparison should also control for differences in mobility,
traffic, and propagation conditions. Moreover, average goodput alone may hide
important differences in lower-tail performance, outage, handover reliability,
admission feasibility, and mobility stability. In O-RAN, controller selection
should therefore consider not only service performance but also computational
cost, training dependence, and interpretability, particularly when multiple
xApps operate on shared RAN parameters \cite{wadud2026aipowered}.
In this work, we investigate whether a lightweight resource- and mobility-aware
heuristic can provide a better overall handover trade-off than a more complex
DDQN policy. We compare standard A3, the proposed heuristic, and DDQN-Multi
through a common xApp-oriented decision interface using paired system-level
simulations with common UE trajectories, traffic assignments, gNB
configurations, and ray-traced received-power inputs.

The principal contributions of this work are as follows:
\begin{itemize}[leftmargin=1.3em,labelsep=0.5em]

    \item We design a lightweight resource- and mobility-aware handover
    heuristic that jointly considers radio quality, target-cell load, PRB
    availability, UE motion, and service class. Its scoring weights are
    optimized offline using NSGA-II on training runs separate from
    evaluation, while mobility-adaptive TTT and context-dependent feasibility
    rules adapt decisions to UE conditions. Once configured, the heuristic
    requires no online learning and retains linear online complexity and
    directly interpretable decisions.

    \item We develop a controlled paired evaluation framework for standard A3,
    the proposed heuristic, and DDQN-Multi under a common handover decision
    interface. All controllers receive identical UE trajectories, traffic
    assignments, gNB configurations, and cached ray-traced received-power
    inputs within each evaluated scenario, thereby controlling for exogenous
    mobility, traffic, and radio variation.

    \item We evaluate the controllers using a multi-KPI methodology extending
    beyond mean goodput to lower-tail goodput, outage, degradation duration,
    handover reliability, ping-pong behaviour, admission performance, energy,
    and fairness. Raw KPIs are complemented by CRITIC--TOPSIS ranking and
    equal-weight sensitivity analysis, while a separate complexity and
    deployability analysis characterizes the trade-off between lightweight
    deterministic control and learned DDQN-based handover.
\end{itemize}
\subsection{Related Work}
\label{sec:related_work}

DDQN-based handover has been widely studied in mobile and heterogeneous
networks. Tan \emph{et al.} \cite{tan2022intelligent} applied DDQN to V2N
handover and compared it with conventional A3, while Wang and Li
\cite{wang2024double} considered DDQN-based handover in dual-connectivity
mmWave heterogeneous networks. More recently, Mondal \emph{et al.}
\cite{mondal2025intelligent} incorporated richer mobility context and explicit
handover and ping-pong costs, and Liang \emph{et al.}
\cite{liang2026speed} introduced speed-adaptive DDQN for high-mobility
heterogeneous networks.
These studies primarily improve the learned handover policy itself. In
contrast, this work examines whether the additional learning complexity is
justified by comparing a multi-feature DDQN with standard A3 and an
offline-optimized, resource- and mobility-aware heuristic under the same
handover interface and matched scenario conditions.

\section{System Model and Handover Controllers}
\label{sec:system_model}

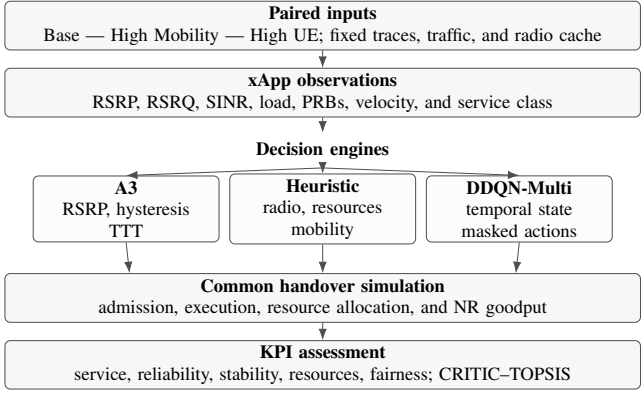
\begin{figure}[t]
\centering
\begin{tikzpicture}[
    font=\scriptsize,
    node distance=2.3mm and 1.3mm,
    >=Latex,
    stage/.style={
        draw=black!70,
        rounded corners=2pt,
        align=center,
        inner xsep=3pt,
        inner ysep=2pt,
        line width=0.35pt,
        fill=black!3
    },
    engine/.style={
        stage,
        text width=0.255\columnwidth,
        minimum height=8mm,
        fill=black!1
    },
    arrow/.style={
        -{Latex[length=1.5mm,width=1.1mm]},
        draw=black!70,
        line width=0.35pt
    }
]
\node[stage, text width=0.93\columnwidth] (inputs)
{\textbf{Paired inputs}\\
Base | High Mobility | High UE; fixed traces, traffic, and radio cache};

\node[stage, below=of inputs, text width=0.93\columnwidth] (obs)
{\textbf{xApp observations}\\
RSRP, RSRQ, SINR, load, PRBs, velocity, and service class};

\node[below=2mm of obs] (controllers)
{\textbf{Decision engines}};

\node[engine, below=1mm of controllers, xshift=-0.295\columnwidth] (a3)
{\textbf{A3}\\
RSRP, hysteresis\\
TTT};

\node[engine, right=1.3mm of a3] (heuristic)
{\textbf{Heuristic}\\
radio, resources\\
mobility};

\node[engine, right=1.3mm of heuristic] (ddqn)
{\textbf{DDQN-Multi}\\
temporal state\\
masked actions};

\node[stage, below=3.8mm of heuristic, text width=0.93\columnwidth] (sim)
{\textbf{Common handover simulation}\\
admission, execution, resource allocation, and NR goodput};

\node[stage, below=of sim, text width=0.93\columnwidth] (eval)
{\textbf{KPI assessment}\\
service, reliability, stability, resources, fairness; CRITIC--TOPSIS};

\draw[arrow] (inputs.south) -- (obs.north);
\draw[arrow] (obs.south) -- (controllers.north);
\draw[arrow] (controllers.south) -- (a3.north);
\draw[arrow] (controllers.south) -- (heuristic.north);
\draw[arrow] (controllers.south) -- (ddqn.north);
\draw[arrow] (a3.south) -- ([xshift=-0.29\columnwidth]sim.north);
\draw[arrow] (heuristic.south) -- (sim.north);
\draw[arrow] (ddqn.south) -- ([xshift=0.29\columnwidth]sim.north);
\draw[arrow] (sim.south) -- (eval.north);
\end{tikzpicture}
\caption{Paired evaluation workflow for A3, the resource-aware heuristic, and
DDQN-Multi under a common handover simulation and KPI-assessment pipeline.}
\label{fig:workflow}
\end{figure}

We model handover control as a near-RT RIC xApp, while keeping the
evaluation independent of a full E2-connected O-RAN deployment. The controller
receives UE- and cell-level observations, including RSRP, RSRQ, SINR, UE
velocity, cell load, PRB availability, service class, and serving/candidate
cell identifiers. It then selects either a target cell or a stay action. This
common xApp-oriented interface allows A3, the resource-aware heuristic, and
DDQN-Multi to be compared as alternative decision engines for the same
mobility-control task \cite{polese2023oran}.
Figure~\ref{fig:workflow} summarizes the paired workflow used to test whether a
simple deterministic heuristic can provide a better handover trade-off than a
more complex learned policy. All controllers use the same admission and handover procedure.
\subsection{Network and Radio Model}

\begin{figure}[t]
\centering
\includegraphics[width=\columnwidth]
{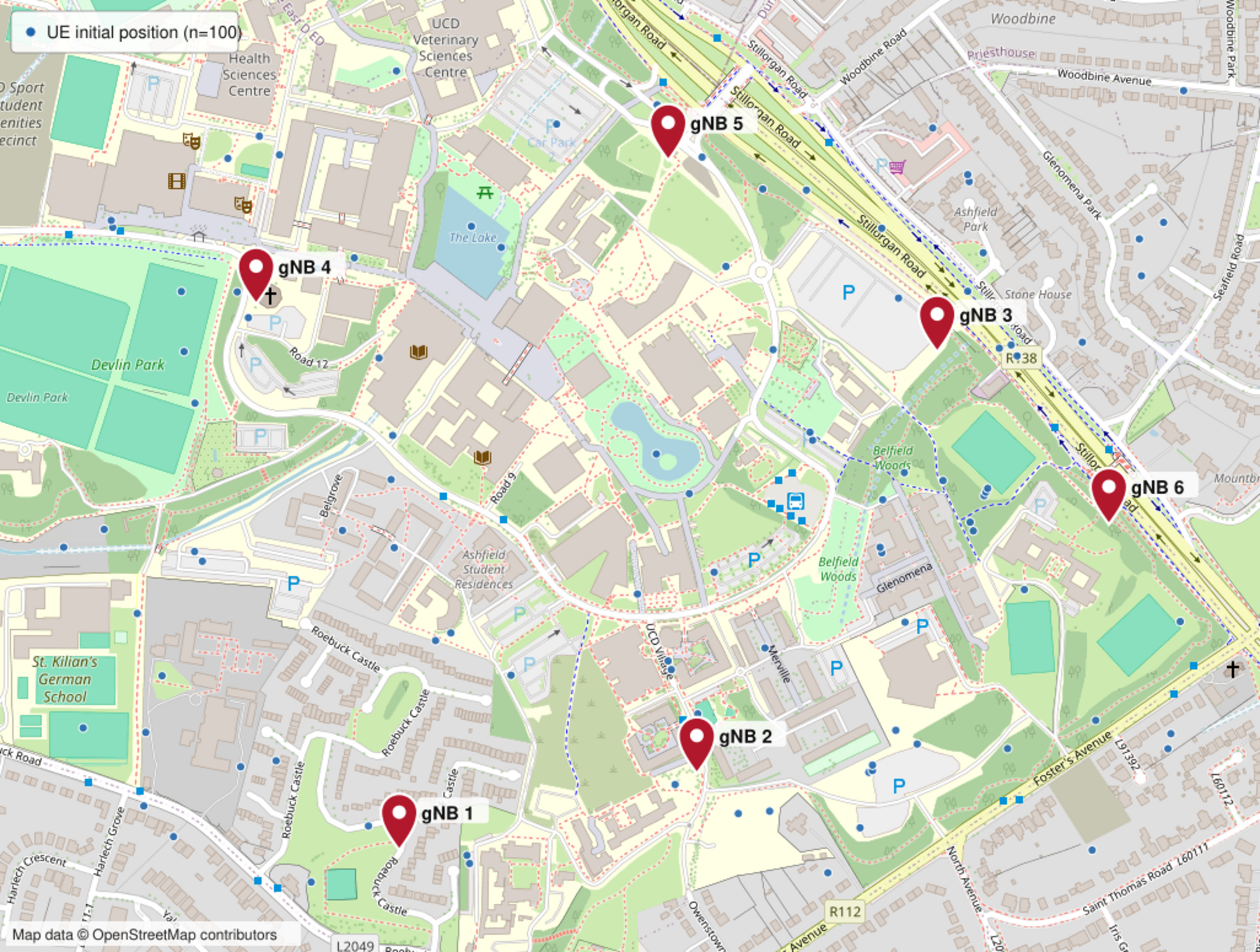}
\caption{The UCD Belfield simulation area, showing the
six simulated NR gNBs and the initial positions of 100 UEs.}\vspace{-10pt}
\label{fig:deployment_map}
\end{figure}

The simulated network covers the campus of University College Dublin Belfield.
Six geographically sourced tower locations are used as NR gNB sites. The gNBs operate at \SI{3.5}{GHz}, with \SI{43}{dBm} conducted transmit power and \SI{25}{m} antenna height. Carrier bandwidth is \SI{100}{MHz} with \SI{30}{kHz} subcarrier spacing and 273 PRBs. Each NR slot of \SI{0.5}{ms} contains 14 OFDM symbols. Mobility, radio conditions, resource allocation and controller states are updated every \SI{20}{ms}, i.e. 40 NR slots.
Figure~\ref{fig:deployment_map} shows the study area, six gNBs, and initial UE
positions.
Directional gNB antenna patterns are defined according to the 3GPP element abstraction with a maximum gain of \SI{8}{dBi}, a horizontal and vertical beamwidth of \SI{65}{\degree}, a maximum attenuation of \SI{30}{dB}, a polarisation slant of $+45^{\circ}$ and an electrical downtilt of \SI{12}{\degree} \cite{3gpp_tr_38901}. Site specific azimuths are used to orient the antenna boresights to the study area. The geographic propagation scene uses local OpenStreetMap building geometry, additional building-height information, automatic ray-tracer building-material assignment and loam terrain.
The shooting-and-bouncing-ray propagation is set up with a small angular separation, up to two reflections and one edge diffraction. These propagation settings are system level modelling assumptions, not parameters imposed by 3GPP.
Let $P_{u,c}(t)$ denote the total ray-traced power received by UE $u$ from
cell $c$ at time $t$. The serving-cell SINR is
computed as

\begin{equation}
\gamma_u(t)=
\frac{P_{u,s_u}(t)}
{N_u+\sum_{c\neq s_u}\rho_c(t)P_{u,c}(t)},
\label{eq:sinr}
\end{equation}

where $s_u$ is the serving cell, $N_u$ is the thermal-noise power including the UE noise figure, and $\rho_c(t)$ is the overlapping PRB activity of the interfering cell. Thus, the effective link quality is affected by both propagation and instantaneous utilisation of resources.

Only controller-independent ray-traced received powers are cached; attachment,
SINR, scheduling, goodput, and handover states are recomputed independently
for each controller. Cache metadata prevents reuse across incompatible
scenarios or radio configurations.

\subsection{Resource Allocation and Model-Derived NR Goodput}

The scheduler dynamically allocates 273 PRBs among connected UEs at each
system-level step. We consider two QoS-lite service classes: eMBB users request
eight PRBs and latency-sensitive users request twelve PRBs. A 10\% cell-level
admission margin is maintained. All controllers use the same traffic,
scheduler, and admission procedure.

The primary KPI is model-derived expected NR goodput. The serving-cell SINR
is mapped to the highest entry in 3GPP TS~38.214 MCS Table~5.1.3.1-1 whose
spectral efficiency does not exceed the efficiency-adjusted Shannon bound
\cite{3gpp_ts_38214}. The model then calculates the transport-block size from
the allocated PRBs, 12 PDSCH symbols per slot, DM-RS overhead, and one
transmission layer. The reported goodput is

\begin{equation}
G_u(t)=
\frac{B^{\mathrm{TBS}}_u(t)}
     {T_{\mathrm{slot}}}
\left(1-\mathrm{BLER}_{\mathrm{target}}\right).
\label{eq:goodput}
\end{equation}

where $T_{\mathrm{slot}}=\SI{0.5}{ms}$ and $\mathrm{BLER}_{\mathrm{target}}=0.1$. This value is calculated exactly within the implemented system-level model,
but represents expected NR goodput at the configured target BLER rather than
measured application throughput or realized HARQ delivery.

\subsection{Standard Radio-Only A3}

The first controller is standard radio-only Event A3. For serving cell $s$
and neighbouring cell $c$, the A3 condition is satisfied when

\begin{equation}
M_{u,c}^{\mathrm{L3}}(t)>
M_{u,s}^{\mathrm{L3}}(t)+H_{\mathrm{A3}},
\label{eq:a3}
\end{equation}

where $M^{\mathrm{L3}}$ is the Layer-3-filtered RSRP and
$H_{\mathrm{A3}}=\SI{3}{dB}$ is the configured hysteresis. The condition must
remain continuously satisfied for a TTT of \SI{640}{ms} before a request is
issued. These parameters follow valid 3GPP measurement-event configuration
choices \cite{3gpp_ts_38331}.

The A3 decision stage uses radio measurements without pre-request PRB
filtering, providing a transparent standardized baseline. Once triggered, the
target proceeds through the common preparation, admission, execution, and
completion process applied to every policy.

\subsection{Resource-Aware Heuristic}
The heuristic controller evaluates the radio, resource, load, and mobility
suitability of each candidate cell. Its nominal score can be expressed as
\begin{equation}
S_{u,c} =
w_R\bar{R}_{u,c}+
w_\gamma\bar{\gamma}_{u,c}+
w_Q\bar{Q}_{u,c}-
w_L\bar{L}_c+
w_P\bar{P}_c,
\label{eq:heuristic}
\end{equation}
where $\bar{R}$, $\bar{\gamma}$, and $\bar{Q}$ are the normalized RSRP,
SINR, and RSRQ, while $\bar{L}$ and $\bar{P}$ denote normalized cell load
and available PRBs. The implementation clips all inputs to $[0,1]$ using
RSRP, SINR, and RSRQ ranges of $[-120,-60]$ dBm, $[-10,30]$ dB, and
$[-20,-10]$ dB, respectively. Cell load is normalized by the configured
load limit, and available PRBs are normalized by the 273-PRB carrier
capacity. Consequently, better radio quality and greater PRB availability
increase the score, whereas higher target-cell load reduces it. The
implemented nominal weights are

\begin{equation}
(w_R,w_\gamma,w_Q,w_L,w_P)=
(0.356,0.374,0.058,0.127,0.087).
\end{equation}

The nominal weights were tuned off-line, in a separate NSGA-II campaign, using Base-scenario training runs separate from the evaluation run reported in this paper. NSGA-II used a population size of 16 over six generations to maximise end-to-end handover completion jointly with minimising final handover failure and ping-pong rate \cite{deb2002nsgaii}. The coefficients selected were then fixed and used as such in all evaluation scenarios.

Nominal weights are adapted to the UE context and renormalised before candidate scoring. High-speed and latency-sensitive UEs get additional SINR weight and eMBB UEs get additional PRB weight. The top-scoring candidate must beat the serving-cell score by an amount that depends on motion and serving quality, satisfy the radio-improvement gates, and pass the target-resource feasibility check. The nominal TTT is \SI{480}{ms} and high-speed UEs use \SI{200}{ms}.

\subsection{DDQN Mobility Policies}

DDQN controller observes the current serving cell and three strongest candidate cells. Each frame contains 64 features: 16 serving/network features and 16 for each
of three candidates. Four consecutive frames form the 256-dimensional state.
The four actions are either to keep the serving cell or to choose one from the three ranked candidates. Radio- and resource-validity masks discard infeasible selections.

The online and target networks have two hidden layers with 128 and 64 fully connected units, respectively. Training is performed using Double Q-learning, prioritised experience replay, a replay buffer of 200,000 transitions, mini-batches of 64, discount factor $\gamma=0.99$, learning rate $10^{-4}$, gradient clipping, and a target-network update every 200 learning steps. 
The exploration rate is decayed from 0.80 to 0.05 during the training.

The reward is driven by successfully delivered NR goodput. Defining
\begin{equation}
U_T(T_u)=\frac{T_u}{T_u+T_{\mathrm{th}}},
\qquad T_{\mathrm{th}}=\SI{1}{Mbps},
\end{equation}

The service reward is positive when the delivered goodput is above the service threshold, and it is a smooth degradation cost below the service threshold. The highest costs are incurred by final handover failures and ping-pong outcomes. The rewards obtained during preparation, execution and completion are stored in the semi-Markov transition and the agent can take into account the full handover interruption and outcome.

DDQN-Multi was trained over eight 60-s episodes, which include two full cycles over Base, Challenging-Radio, High-Mobility, and High-UE conditions.
Training seeds were 201–204 and 211–214. For held-out seed 9001 (excluded from training), the checkpoints saved at the end of each complete cycle were evaluated over the four conditions. For these held-out evaluations we chose the checkpoint with the best mean NR goodput. The final paired controller evaluation was performed with seed 1001, independent of the training and checkpoint-validation seeds.

\section{Evaluation Methodology}
\label{sec:evaluation}
\begin{figure*}[t]
\centering
\includegraphics[
    width=0.99\textwidth,
    trim=0cm 0cm 0cm 0cm,
    clip
]{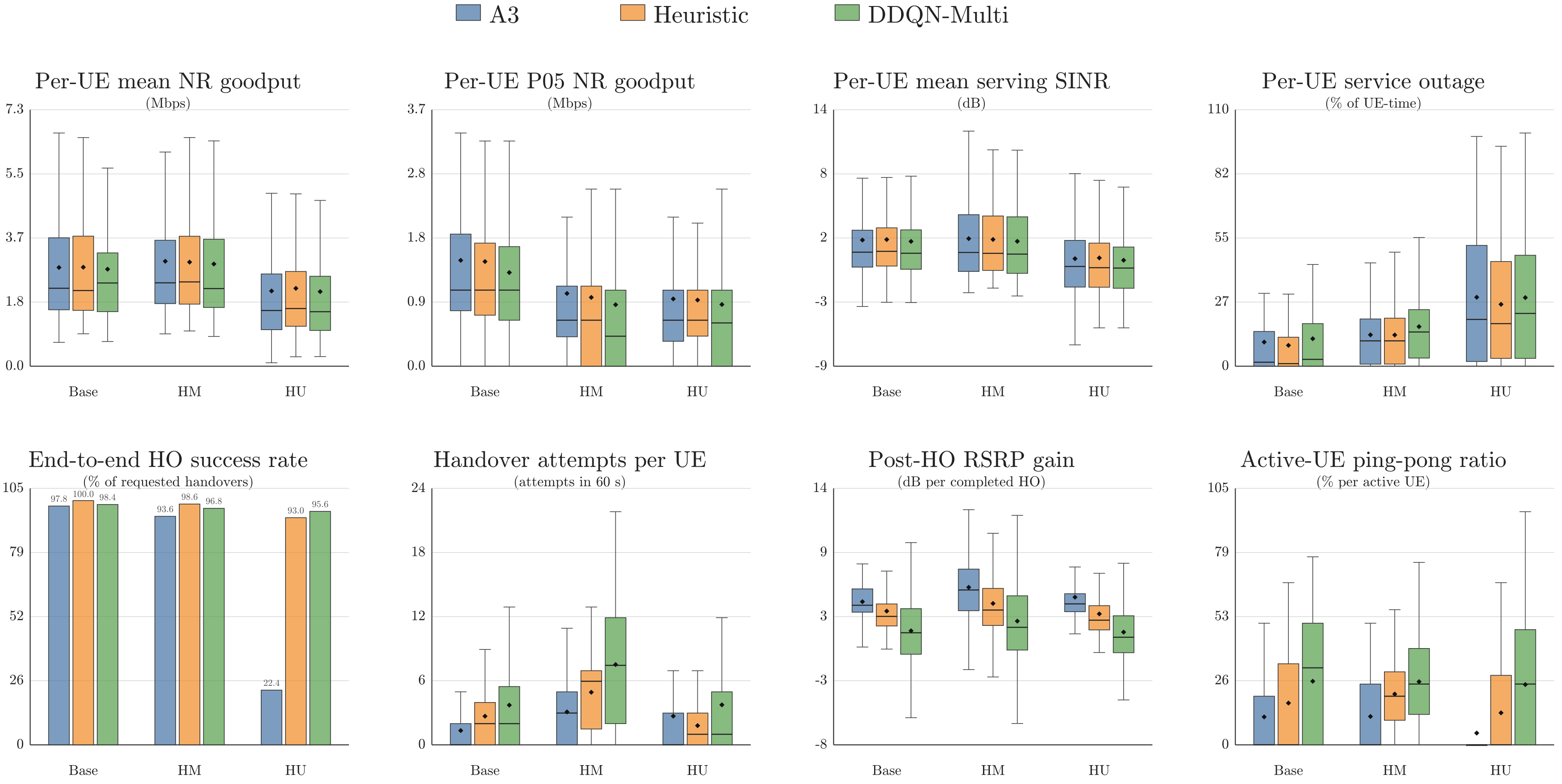}
\caption{Sample-level KPI distributions across the paired 60-s evaluation.}
\label{fig:kpi_boxplots}
\end{figure*}

\begin{table}[t]
\centering
\caption{Controller performance and scenario-specific TOPSIS scores.}
\label{tab:combined_results}

\scriptsize
\setlength{\tabcolsep}{1.0pt}
\renewcommand{\arraystretch}{0.95}

\setlength{\aboverulesep}{0pt}
\setlength{\belowrulesep}{0pt}

\resizebox{\columnwidth}{!}{%
\begin{tabular}{@{}l*{10}{c}@{}}
\toprule
Scen. & Ctrl. &
\shortstack{TP\\(Mbps)} &
\shortstack{P05\\(Mbps)} &
\shortstack{Out.\\\tiny{\%}} &
\shortstack{Succ.\\HO} &
\shortstack{PP\\\tiny{\%}} &
\shortstack{PRB\\Fail.\tiny{\%}} &
\shortstack{CRITIC\\TOPSIS} &
\shortstack{Equal\\TOPSIS} &
Rank \\
\midrule

\multirow{3}{*}{Base}
& A3
& 2.827
& 0.734
& 10.34
& 132/135
& 25.76
& 2.22
& 0.458
& 0.606
& 2 \\

& \textbf{Heuristic}
& \textbf{2.834}
& \textbf{0.864}
& \textbf{8.93}
& 270/270
& 30.74
& \textbf{0.00}
& \textbf{0.775}
& \textbf{0.786}
& \textbf{1} \\

& DDQN-M.
& 2.778
& 0.734
& 11.77
& 369/375
& 36.04
& 0.27
& 0.138
& 0.073
& 3 \\

\midrule

\multirow{3}{*}{\shortstack[l]{High\\Mob.}}
& A3
& \textbf{3.005}
& 0.576
& 13.43
& 291/311
& \textbf{14.09}
& 5.79
& 0.320
& \textbf{0.748}
& 3 \\

& \textbf{Heuristic}
& 2.980
& \textbf{0.576}
& \textbf{13.38}
& 489/496
& 22.49
& \textbf{0.00}
& \textbf{0.865}
& 0.711
& \textbf{1} \\

& DDQN-M.
& 2.927
& 0.432
& 16.94
& 734/758
& 28.61
& 0.53
& 0.506
& 0.157
& 2 \\

\midrule

\multirow{3}{*}{\shortstack[l]{High\\UE}}
& A3
& 2.154
& 0.000
& 29.52
& 91/406
& \textbf{9.89}
& 77.59
& 0.481
& 0.405
& 2 \\

& \textbf{Heuristic}
& \textbf{2.229}
& \textbf{0.432}
& \textbf{26.49}
& 253/272
& 27.67
& 6.25
& \textbf{0.704}
& \textbf{0.789}
& \textbf{1} \\

& DDQN-M.
& 2.135
& 0.000
& 29.38
& 541/566
& 39.00
& \textbf{3.36}
& 0.419
& 0.286
& 3 \\

\bottomrule
\end{tabular}%
}
\end{table}
\subsection{Scenarios and Paired Experimental Design}

The evaluation considers three system-level simulation scenarios. In the \emph{Base} scenario 100 UEs are moving according to a random-waypoint model with speeds between \SIrange{0.5}{2}{m/s}. The \emph{High-Mobility} scenario comprises 100 UEs moving at \SIrange{8}{12}{m/s} with non-synchronized changes in direction, thus increasing the number of boundary crossings and decreasing the residence time in a cell. The \emph{High-UE} scenario scales the number of UEs to 150, while the normal mobility model and per-UE offered PRB demand are retained, thus creating a resource-contention stress condition.

Each simulation is run for \SI{60}{s} with 3,000 system-level updates.
Twenty percent of UEs are assigned to a controlled mobility-stress cohort that rotates around known handover-prone locations. All the policies have the same cohort assignment. This allows for repeatable exposure of cell-edge transitions while the other UEs assume the general mobility distribution of the scenario.


\subsection{Performance Indicators}

We evaluate service, radio, mobility, reliability, resource use, and fairness. The main service metrics are mean and fifth-percentile NR goodput, total service outage, 95th-percentile service-degradation duration, and the QoS-satisfaction ratio. Radio performance is defined by serving-cell RSRP, RSRQ, SINR and immediate post-handover radio gain.

A handover attempt is successful if it completes the preparation, execution, completion and path switching. Its end-to-end success rate is the number of completions divided by the total number attempted. Mobility stability is quantified by considering a successful-handover interruption, attempts per UE-minute, and ping-pongs per completed handover, where a ping-pong is a return to a recently serving cell within the configured observation interval. Resource efficiency is evaluated by PRB satisfaction, target-admission failures, cell-load balance and energy per delivered Mbit. User-level goodput fairness is evaluated using Jain's fairness index based on
the time-averaged goodput delivered to each UE \cite{jain1984fairness}.

\subsection{CRITIC--TOPSIS Multi-KPI Ranking}

Mobility policies can generate different trade-offs and, therefore, the raw KPIs are accompanied by a CRITIC–TOPSIS ranking. There are fourteen indicators:
mean and fifth percentile goodput, total outage, 95th percentile degradation duration, QoS satisfaction, end-to-end handover success, successful handover interruption, ping-pong ratio, handover attempts per UE-minute, post-handover RSRP gain, PRB satisfaction, energy per delivered Mbit, Jain fairness, and orphan/RLF events per UE-minute.

For controller $i$ and KPI $j$, min--max normalization produces a
benefit-oriented value $z_{ij}$: $(x_{ij}-x_j^{\min})/(x_j^{\max}-x_j^{\min})$
for benefit KPIs and $(x_j^{\max}-x_{ij})/(x_j^{\max}-x_j^{\min})$ for cost
KPIs. CRITIC assigns greater importance to KPIs that provide both contrast
among controllers and non-redundant information \cite{diakoulaki1995critic}:
\begin{equation}
C_j=\sigma_j\sum_{k=1}^{m}(1-\rho_{jk}),
\qquad
w_j=\frac{C_j}{\sum_{\ell=1}^{m}C_\ell},
\label{eq:critic}
\end{equation}

where $\sigma_j$ and $\rho_{jk}$ are the normalized standard deviation and
inter-KPI correlation, respectively.

The weighted normalized matrix is $v_{ij}=w_jz_{ij}$. TOPSIS then evaluates
the distance of every controller from the empirical ideal and anti-ideal
solutions \cite{hwang1981topsis}:

\begin{equation}
\begin{aligned}
D_i^{\pm}&=\sqrt{\sum_j(v_{ij}-v_j^{\pm})^2},\\
S_i&=\frac{D_i^{-}}{D_i^{+}+D_i^{-}},\qquad 0\leq S_i\leq1.
\end{aligned}
\label{eq:topsis_score}
\end{equation}

where a larger value indicates a policy closer to the ideal multi-KPI
operating point.

CRITIC--TOPSIS is accompanied by equal-weight TOPSIS to examine sensitivity
to the data-derived weights. Because only three primary controllers and one
paired evaluation are reported, the scenario-specific scores are interpreted
as descriptive summaries of the observed KPI trade-offs. They are not treated
as confidence intervals, statistical significance tests, or universal
controller rankings. Multi-seed paired bootstrap analysis is reserved for an
extended evaluation.

\section{Results and Discussion}
\label{sec:results}

\subsection{Overall Performance}

Table~\ref{tab:combined_results} compares A3, the heuristic, and DDQN-Multi using
the same UE and radio inputs within each scenario.

\subsection{Base Scenario}

Figure~\ref{fig:kpi_boxplots} shows that the controllers produce broadly
similar mean serving SINR and mean goodput, but substantially different mobility
behaviour. The heuristic achieves the highest P05 goodput
(\SI{0.864}{Mbps}) and lowest outage (8.93\%) while avoiding target-admission
failures. A3 performs fewer handovers and generally obtains larger post-HO RSRP
gains, consistent with its conservative radio-only trigger. DDQN-Multi instead
initiates more handovers and exhibits the highest ping-pong activity without a
corresponding goodput improvement. The heuristic therefore occupies an
intermediate operating point: it reacts more actively than A3 but avoids much
of the mobility churn observed with DDQN-Multi.

\subsection{High-Mobility Scenario}

High mobility highlights a different trade-off. A3 achieves the highest mean
goodput (\SI{3.005}{Mbps}) and lowest ping-pong ratio, while the heuristic
achieves nearly identical mean goodput (\SI{2.980}{Mbps}) and outage
(13.38\% versus 13.43\%) with higher handover completion and no admission
failures. DDQN-Multi generates substantially more handover attempts and
ping-pong events, while obtaining lower goodput and higher outage. Thus, when
mobility rather than resource congestion dominates, conservative A3 remains a
strong operating point, while resource awareness primarily improves target
feasibility rather than radio performance.

\subsection{High-UE Scenario}

The High-UE scenario exposes the effect of resource awareness most clearly.
Although A3 continues to select targets with relatively large post-HO RSRP
gains, 77.59\% of its handover requests fail target admission, reducing its
end-to-end completion rate to 22.41\%. The heuristic and DDQN-Multi reduce
admission failure to 6.25\% and 3.36\%, respectively. However, DDQN-Multi
requires 566 attempts and exhibits the highest ping-pong activity, whereas the
heuristic requires only 272 attempts while achieving 93.01\% completion.
Consequently, the heuristic obtains the lowest outage (26.49\%) and the only
non-zero P05 goodput (\SI{0.432}{Mbps}). These results indicate that under
congestion, target feasibility is important, but additional handover activity
does not necessarily translate into better service continuity.

\subsection{Multi-KPI Ranking}

Table~\ref{tab:combined_results} reports the scenario-specific multi-KPI scores.
Scores are normalized independently within each scenario and are therefore
not compared or averaged across scenarios.

The heuristic ranks first under both weighting methods in Base and High UE.
In High Mobility, CRITIC favours the heuristic's reliability and admission
performance, whereas equal weighting favours A3's goodput and stability.
This difference illustrates the sensitivity of composite rankings to KPI
weighting and reinforces the need to interpret the scores together with the
underlying raw KPIs.
\subsection{Computational Complexity and Deployability}
\label{subsec:complexity}

A3 is the least complex in online evaluations since the relative radio
measurements are evaluated and the TTT state is maintained. Also the designed
heuristic evaluates radio quality, resources, mobility and service class but
its complexity is still $O(C)$ since the feature set is fixed. DDQN-Multi
requires a $256$-dimensional temporal state, action masking and a
$256$--$128$--$64$--$4$ neural-network forward pass, which is $41,216$
multiply--accumulates per UE decision, excluding feature construction and
masking. This comparison, along with the observed KPI results, places the
heuristic as a lightweight and interpretable operating point between
radio-only A3 and learned DDQN control. The trade-off is summarised in
Table~\ref{tab:complexity}.

\begin{table*}[ht]
\centering
\caption{Controller complexity, offline preparation, and deployment
characteristics.}
\label{tab:complexity}
\scriptsize
\setlength{\tabcolsep}{2pt}
\begin{tabular}{
p{0.11\textwidth}
p{0.23\textwidth}
p{0.19\textwidth}
p{0.21\textwidth}
p{0.19\textwidth}}
\toprule
Controller &
Decision information &
Online computation per UE &
Offline preparation &
Deployment characteristics \\
\midrule

A3 &
Serving- and candidate-cell RSRP, hysteresis, and TTT &
$O(C)$; no learned model parameters &
None; hysteresis and TTT are configured before operation &
Lowest complexity and direct interpretation of every decision \\

\midrule

Resource-aware heuristic &
RSRP, RSRQ, SINR, cell load, available PRBs, UE mobility, and service class &
$O(CF)$, equivalent to $O(C)$ for the fixed feature set &
Five score weights tuned by NSGA-II using a population of 16 over six generations; coefficients are fixed during evaluation &
Lightweight execution without a neural-network runtime; decision scores and feasibility gates remain interpretable \\

\midrule

DDQN-Multi &
Four stacked 64-feature frames, producing a 256-dimensional state; three candidate cells and an action-validity mask &
41,412 trainable parameters and 41,216 multiply--accumulates per forward pass, excluding activation, feature-construction, and masking costs &
Eight 60-s training episodes followed by checkpoint evaluation using held-out seed 9001 &
Requires a trained checkpoint and neural-network inference; supports temporal decisions but has lower direct interpretability \\

\bottomrule
\end{tabular}

\vspace{1mm}
\parbox{0.97\textwidth}{\scriptsize
$C$ denotes the number of candidate cells and $F$ the fixed number of
features evaluated per candidate. All controllers operate at the common
20-ms system-level decision interval.}
\end{table*}
\section{Conclusion}
\label{sec:conclusion}

In this paper, we design and evaluate a resource-aware heuristic and a service-aware DDQN mobility policy for an O-RAN xApp, taking radio-only 3GPP Event A3 as the reference baseline. A paired system-level framework ensured that each controller saw the same UE trajectories, traffic assignments, and cached ray-traced radio conditions.

The results show that similar average goodput can hide large differences in
outage, handover reliability, admission feasibility and mobility stability.
The designed resource-aware heuristic achieves the best overall KPI balance
and lowest outage in all three scenarios in the reported paired evaluation
whilst maintaining linear online complexity, fixed coefficients and directly
traceable decisions. A3 is still the simplest controller, but it is not
resource-aware under congestion; DDQN-Multi has strong target feasibility but
generates more handover activities, is training-dependent and computationally
complex. The results thus indicate lightweight resource-aware control as a
promising candidate for mobility-management xApps. Future work will extend the
evaluation to multiple paired seeds and an E2-connected near-RT RIC.
\section{acknowledgments}
Funding for this research is provided by the European Union's Horizon Europe research and innovation program through the Marie Skłodowska-Curie SE grant under the agreement RE-ROUTE No 101086343 and Research Ireland CONNECT-2 grant 13/RC/2077-P2.
\bibliographystyle{IEEEtran}
\bibliography{references}

@techreport{3gpp_tr_38901,
  author       = {{3GPP}},
  title        = {{Study on channel model for frequencies from 0.5 to 100 GHz}},
  institution  = {3rd Generation Partnership Project},
  type         = {TR},
  number       = {38.901},
  version      = {19.2.0},
  note         = {Release 19},
  year         = {2025}
}

@article{wadud2026aipowered,
  author = {Abdul Wadud and Fatemeh Golpayegani and Nima Afraz},
  title = {{AI}-Powered Conflict Management in {Open RAN}: Detection, Classification, and Mitigation},
  journal = {Computer Networks},
  volume = {286},
  pages = {112432},
  year = {2026},
  doi = {10.1016/j.comnet.2026.112432}
}

@article{tan2022intelligent,
  author  = {Kang Tan and Duncan Bremner and Julien Le Kernec and
             Yusuf A. Sambo and Lei Zhang and Muhammad Ali Imran},
  title   = {Intelligent Handover Algorithm for Vehicle-to-Network
             Communications With Double-Deep Q-Learning},
  journal = {IEEE Transactions on Vehicular Technology},
  volume  = {71},
  number  = {7},
  pages   = {7848--7862},
  year    = {2022},
  doi     = {10.1109/TVT.2022.3169804}
}

@article{mondal2025intelligent,
  author  = {Sudeb Mondal and Sawera Aslam and Daud Khan and Kyunghi Chang},
  title   = {Intelligent Handover Orchestration in Beyond {5G} and
             Urban {V2X} Dual Connectivity Networks: A Deep Reinforcement
             Learning Approach},
  journal = {IEEE Access},
  volume  = {13},
  pages   = {196509--196525},
  year    = {2025},
  doi     = {10.1109/ACCESS.2025.3633314}
}

@article{liang2026speed,
  author  = {Gen Liang and Fengyi Su and Baiyang Lu and Hewei Yu},
  title   = {A Speed-Adaptive Handover Decision Algorithm for Ultra-Dense
             Heterogeneous Wireless Networks},
  journal = {Ad Hoc Networks},
  volume  = {192},
  pages   = {104313},
  year    = {2026},
  doi     = {10.1016/j.adhoc.2026.104313}
}

@article{wang2024double,
  author  = {Hao Wang and Bo Li},
  title   = {Double-Deep Q-Learning-Based Handover Management in
             mmWave Heterogeneous Networks With Dual Connectivity},
  journal = {Transactions on Emerging Telecommunications Technologies},
  volume  = {35},
  number  = {1},
  pages   = {e4907},
  year    = {2024},
  doi     = {10.1002/ett.4907}
}

@techreport{3gpp_ts_38331,
  author       = {{3GPP}},
  title        = {{NR; Radio Resource Control (RRC); Protocol specification}},
  institution  = {3rd Generation Partnership Project},
  type         = {TS},
  number       = {38.331},
  note         = {Release 18},
  year         = {2025}
}

@article{haghrah2023handover_nr,
  author = {Amiraslan Haghrah and Mehran Pourmohammad Abdollahi and Hosein Azarhava and Javad Musevi Niya},
  title = {A Survey on the Handover Management in {5G-NR} Cellular Networks: Aspects, Approaches and Challenges},
  journal = {EURASIP Journal on Wireless Communications and Networking},
  volume = {2023},
  number = {52},
  year = {2023},
  doi = {10.1186/s13638-023-02261-4}
}

@techreport{3gpp_ts_38214,
  author       = {{3GPP}},
  title        = {{NR; Physical layer procedures for data}},
  institution  = {3rd Generation Partnership Project},
  type         = {TS},
  number       = {38.214},
  version      = {19.2.0},
  note         = {Release 19},
  year         = {2025}
}

@article{palas2021multicriteria,
  author  = {Palas, M. R. and Islam, M. R. and Roy, P. and Razzaque, M. A. and Alsanad, A. and AlQahtani, S. A. and Hassan, M. M.},
  title   = {{Multi-criteria handover mobility management in 5G cellular network}},
  journal = {Computer Communications},
  volume  = {174},
  pages   = {81--91},
  month   = jun,
  year    = {2021},
  doi     = {10.1016/j.comcom.2021.04.020}
}

@techreport{jain1984fairness,
  author      = {Jain, Raj and Chiu, Dah-Ming and Hawe, William R.},
  title       = {{A quantitative measure of fairness and discrimination for resource allocation in shared computer systems}},
  institution = {Digital Equipment Corporation},
  address     = {Hudson, MA, USA},
  number      = {DEC-TR-301},
  year        = {1984}
}

@article{diakoulaki1995critic,
  author  = {Diakoulaki, D. and Mavrotas, G. and Papayannakis, L.},
  title   = {{Determining objective weights in multiple criteria problems: The CRITIC method}},
  journal = {European Journal of Operational Research},
  volume  = {55},
  number  = {2},
  pages   = {763--770},
  year    = {1995}
}

@book{hwang1981topsis,
  author    = {Hwang, Ching-Lai and Yoon, Kwangsun},
  title     = {{Multiple Attribute Decision Making: Methods and Applications}},
  publisher = {Springer-Verlag},
  address   = {Berlin, Germany},
  year      = {1981}
}

@article{deb2002nsgaii,
  author  = {Deb, Kalyanmoy and Pratap, Amrit and Agarwal, Sameer and Meyarivan, T.},
  title   = {{A fast and elitist multiobjective genetic algorithm: NSGA-II}},
  journal = {IEEE Transactions on Evolutionary Computation},
  volume  = {6},
  number  = {2},
  pages   = {182--197},
  month   = apr,
  year    = {2002},
  doi     = {10.1109/4235.996017}
}

@article{polese2023oran,
  author  = {Polese, Michele and Bonati, Leonardo and D'Oro, Salvatore and Basagni, Stefano and Melodia, Tommaso},
  title   = {{Understanding O-RAN: Architecture, interfaces, algorithms, security, and research challenges}},
  journal = {IEEE Communications Surveys \& Tutorials},
  volume  = {25},
  number  = {2},
  pages   = {1376--1411},
  year    = {2023},
  doi     = {10.1109/COMST.2023.3239220}
}

\end{document}